\documentclass[aps,prl,twocolumn,superscriptaddress]{revtex4}

\usepackage{graphicx}
\usepackage{color}
\usepackage{changes} %allows for \deleted{}
\usepackage{xfrac} %allows \sfrac{1}{3} = 1/3
\usepackage[hidelinks]{hyperref}
\usepackage{gensymb}
\begin{document}

\title{Epitaxial single layer NbS$_{2}$ on Au(111): synthesis, structure, and electronic properties}
%different title1: Crystalline and electronic structure of epitaxial single-layer NbS$_{2}$ on Au(111)}}
%different title2: The synthesis, structure and electronic properties of epitaxial single-layer NbS$_{2}$ on Au(111)
\author{Raluca-Maria Stan}
\affiliation{Department of Physics and Astronomy, Interdisciplinary Nanoscience Center (iNANO), Aarhus University, 8000 Aarhus C, Denmark.}
\author{Sanjoy K. Mahatha}
\affiliation{Department of Physics and Astronomy, Interdisciplinary Nanoscience Center (iNANO), Aarhus University, 8000 Aarhus C, Denmark.}
\author{Marco Bianchi}
\affiliation{Department of Physics and Astronomy, Interdisciplinary Nanoscience Center (iNANO), Aarhus University, 8000 Aarhus C, Denmark.}
\author{Charlotte E. Sanders}
\affiliation{Central Laser Facility, STFC Rutherford Appleton Laboratory, Harwell OX11 0QX, United Kingdom}
\author{Davide Curcio}
\affiliation{Department of Physics and Astronomy, Interdisciplinary Nanoscience Center (iNANO), Aarhus University, 8000 Aarhus C, Denmark.}
\author{Philip Hofmann}
	\email{philip@phys.au.dk}
	\affiliation{Department of Physics and Astronomy, Interdisciplinary Nanoscience Center (iNANO), Aarhus University, 8000 Aarhus C, Denmark.}
\author{Jill A. Miwa}
	\email{miwa@phys.au.dk}
\affiliation{Department of Physics and Astronomy, Interdisciplinary Nanoscience Center (iNANO), Aarhus University, 8000 Aarhus C, Denmark.}
	\date{\today}

\begin{abstract}
We report on the epitaxial growth of single layer NbS$_2$ on Au(111) and determine both its crystalline and electronic structure by a combination of low-energy electron diffraction, scanning tunnelling microscopy and angle-resolved photoemission spectroscopy. The layer is found to grow in the 1H structural phase with a lattice constant of (3.29$\pm$0.03)~\AA, a value comparable to the bulk 2H NbS$_2$ lattice constant.  The photoemission data reveals a metallic band structure down to a temperature of 30~K. The observed bands are rather broad and consistent with either a strong NbS$_2$-substrate interaction or with the recently reported interplay of strong many-body effects in single layer NbS$_2$ \cite{Loon:2018aa}. No indications of a charge density wave are observed. 

%The layer is found to be electron doped with a carrier concentration of (3.6$\pm$0.2)$\times$10$^{14}$~cm$^{-1}$. 
\end{abstract}
\maketitle

\section{Introduction}

%\mr{In red I write things that you either need to check (I just wrote what I though was likely without being able to check) or things that need to be fixed depending on the results of further analysis.}\mc{Comments in blue.}\mg{Raluca's writing in green.}
%(metallic) single layer TMDCs
% introduce single layer (SL) and TMDC
Metallic transition metal dichalcogenides (TMDCs) have been studied for decades due to their susceptibility to many fascinating quantum states, such as charge and spin ordering, Mott insulating states, and superconductivity; and because, more generally, they provide a platform for the study of many-body interactions in nearly two-dimensional (2D) materials \cite{Wilson:1969aa,Wilson:1975aa,Rossnagel:2011aa}. The ability to isolate single layer (SL) TMDCs \cite{Mak:2010aa,Novoselov:2005ab,Splendiani:2010aa} now extends these types of investigations to genuinely 2D  systems. This avenue of research is rather compelling, especially since it has turned out that the assumption of a purely 2D character in van der Waals bonded bulk compounds may not be generally valid, after all \cite{Di-Sante:2017aa,Ngankeu:2017aa}. Moreover, the properties of SL materials may be tuned by the choice of substrate, as this interaction can have a significant impact on the electronic properties of the 2D material, for instance, by screening the otherwise strongly enhanced Coulomb interaction \cite{Ugeda:2014aa,Qiu:2013aa,Antonija-Grubisic-Cabo:2015aa,Rosner:2016aa}. Studies of metallic SL TMDCs are considerably more challenging  than of their semiconducting counterparts, such as MoS$_2$, due to their strongly enhanced chemical reactivity, which may also be responsible for the partly contradictory results on, for example, the superconducting transition temperature in atomically thin 2H TaS$_2$ \cite{Galvis:2014aa,Navarro-Moratalla:2016aa,Peng:2018aa} or the charge density wave (CDW) transition temperature in SL NbSe$_2$ \cite{Xi:2015aa,Ugeda:2016aa}.

%What is especially interesting about NbS2?
The isoelectronic sulphides and selenides of vandium, niobium and tantalum show an intriguing competition between magnetism, charge density waves and Mott insulating behaviour. It is not unusual for several of these quantum states to be observed in the same material at different temperatures or even simultaneously. 2H NbS$_2$ stands out in this class of materials because it is widely considered to be the only case for which superconductivity can be observed ($T_c\approx$6~K) in the absence of charge density wave states \cite{Naito:1982aa,Guillamon:2008aa,Heil:2017aa} (even though this long-held point of view has recently been called into question \cite{Leroux:2018aa}). 

While few layer (3R) \cite{Kozhakhmetov:2018aa} and even single layer  NbS$_2$ \cite{Bark:2018aa} have been successfully synthesized, little is known about the different ground states in SL NbS$_2$. Density functional theory and GW calculations predict SL NbS$_2$ to be metallic with a half-filled upper valence band that is well-separated from other electronic states \cite{Kuc:2011aa,Heil:2018aa}. In this approximation, the ground state of the system is not magnetic but it is close to magnetic phases \cite{Zhou:2012ab,Guller:2016aa}. Very recently, the electronic structure of SL NbS$_2$ was studied with the inclusion of several types of many-body effects, such as short- and long-range Coulomb interactions, and electron-phonon interactions \cite{Loon:2018aa}. The study found that each of these interactions has a significant effect on the spectral function, leading to situations that are drastically different from the bare single particle dispersion. Rather unexpectedly, the combined effect of the many-body interactions restores the spectral function to what essentially resembles a broadened version of the bare dispersion. From these results, SL NbS$_2$ might be expected to be a unique case for experimentally tuning different many-body interactions by, for example, the choice of doping or substrate, in order to reach a multitude of desired ground states.

% What we do here:
In this Article, we report on the synthesis of SL NbS$_2$ on Au(111). Due to the epitaxial character of the system, the growth is restricted to a single orientation (and its mirror domain), thereby allowing both the crystalline and electronic structure of the SL NbS$_2$ to be probed with spatially averaging techniques such as low-energy electron diffraction (LEED) and angle-resolved photoemission spectroscopy (ARPES). Consistent with a strong NbS$_2$-Au(111) interaction, or with an interplay of strong many-body effects in the SL NbS$_2$ \cite{Loon:2018aa}, or with a combination of both, the ARPES data divulges a metallic band structure with broad bands.
\section{Experimental}

The growth of epitaxial SL NbS$_{2}$ on Au (111)  was based on a growth technique previously developed for other SL TMDCs \cite{Lauritsen:2007aa,Fuchtbauer:2014aa,Miwa:2015aa,Dendzik:2015aa,Sanders:2016aa,Arnold:2018ab}, with the transition metal being evaporated onto a single-crystal in a sulphur-rich atmosphere. Initially the Au(111) substrate (Mateck) was cleaned by repeated cycles of ion bombardment (Ne$^{+}$ at 1.5\,keV) and annealing to 650~\textdegree C under ultrahigh vacuum  (UHV) conditions. The surface order and cleanliness were established by scanning tunnelling microscopy (STM); the substrate was sputtered and annealed until STM revealed an atomically clean surface with the characteristic herringbone reconstruction \cite{Barth:1990aa}. Niobium was evaporated from an electron beam evaporator (Nb rod of 99.9$\%$ purity from Goodfellow)  onto the clean Au(111) surface, held at room temperature, in a pressure of 2$\times$10$^{-6}$~mbar of dimethyl disulphide  (DMDS, C$_2$H$_6$S$_2$, with $>$99\,\% purity from Sigma-Aldrich). %We note that a similar growth procedure involving H$_2$S gas instead of DMDS is often used to prepare SL TMDCs. Despite the additional sulphur atom in DMDS compared with H$_2$S, both compounds have a similar sulphidation potential \cite{Fuchtbauer:2014aa,Arnold:2018ab} and thus we expect to obtain very similar results using H$_2$S. 
The sample was subsequently annealed for 30~min at $\approx$450~\textdegree C while the same pressure of DMDS was maintained. Finally, the DMDS was pumped out of the chamber and the sample was annealed to $\approx$600~\textdegree C for 30~min in order to eliminate residual physisorbed DMDS, and to improve the overall crystalline quality of the SL NbS$_{2}$. 

Synthesis and STM characterization of the SL NbS$_{2}$ growth were  performed in a dedicated growth chamber (base pressure of 10$^{-10}$\,mbar) using a home-built Aarhus-type STM  \cite{Besenbacher:1988ab}. Tips were made from a Pt-Ir alloy. All STM images were processed using free WSxM software \cite{Horcas:2007aa} and generally involved plane flattening and calibrating to reflect the known atomic lattice parameter of the clean Au(111) surface. The uncertainties associated with the quoted lengths determined by STM arise mainly from thermal drift and/or piezo creep.  LEED and ARPES measurements  were performed at the SGM3 beamline at the ASTRID2 synchrotron radiation facility \cite{Hoffmann:2004aa}. ARPES and LEED data were collected at 30\,K. All ARPES measurements presented here were performed with a photon energy of 50\,eV with an energy and angular resolution better than 25\,meV and 0.2\,\textdegree, respectively.

To transfer the reactive SL NbS$_{2}$, in air, from the growth/STM chamber to the ARPES/LEED chamber, the sample was protected by a DMDS capping layer.  The capping layer was prepared by keeping the grown SL NbS$_{2}$ sample at room temperature in a DMDS pressure of 2$\times$10$^{-6}$~mbar for 30~min.  Once the sample was reintroduced to UHV conditions at the SGM3 beamline, it was annealed to $\approx$600~\textdegree C for 30 min to remove the physisorbed capping layer. STM images were acquired after the removal of the capping layer in the ARPES chamber (not shown), and were consistent with a  surface cleanliness  on par with SL NbS$_{2}$ samples grown without a capping layer.

\section{Results and Discussion}

%start with STM
\begin{figure}
	\begin{center}
		\includegraphics[width=1.0\linewidth]{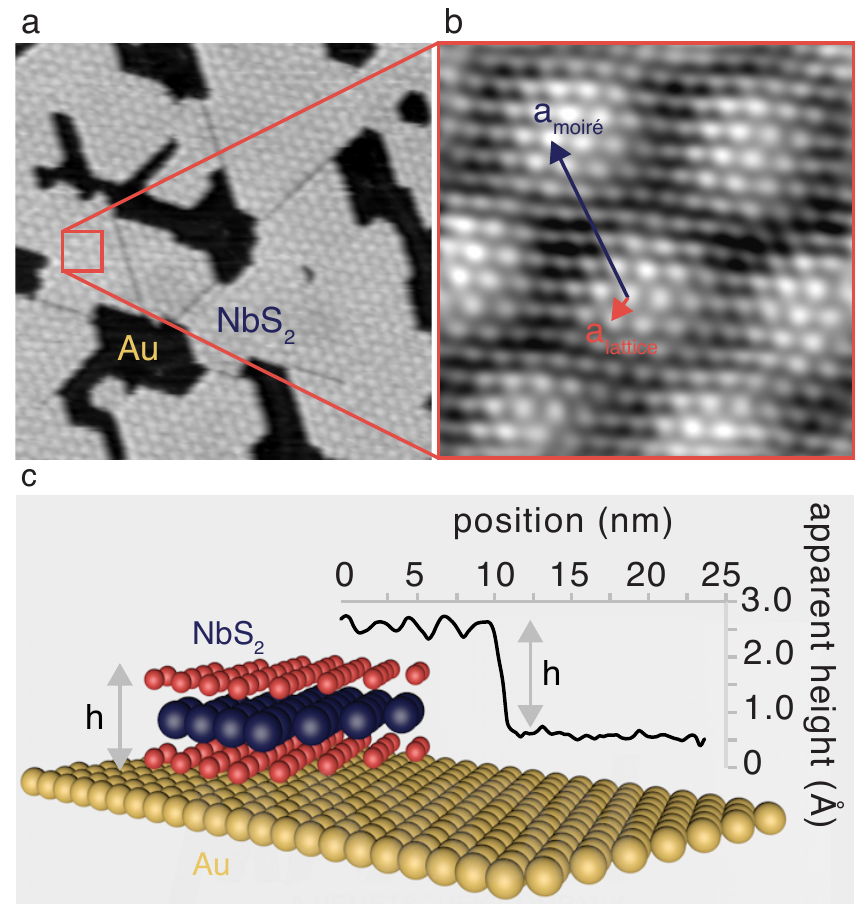}
		\caption{STM characterization of SL NbS$_{2}$ on Au (111). (a) A (621$\times$621)\,{\AA}$^2$ overview revealing irregularly shaped domains and a hexagonal moir\'{e} pattern.  Image parameters:  I$_t$=0.46~nA, V$_b$=-0.86~V. (b) A  (50$\times$50)\,{\AA}$^2$ atomically resolved STM image of the SL NbS$_{2}$. The smaller hexagonal atomic structure and the larger hexagonal moir\'{e} pattern are both visible in the image. Image parameters: I$_t$=1.68~nA, V$_b$=-0.31~V. (c) A schematic of the SL NbS$_{2}$ (sulphur atoms are indicated by red spheres and the niobium by blue ones) on a Au substrate, with the height of the single layer indicated by ``h". A line profile taken across the edge of a NbS$_{2}$ domain with the corresponding measured apparent height demarcated by the double headed grey arrow. The apparent height was measured to be (3.1$\pm$0.2)~{\AA}.} 
		\label{fig:1}
	\end{center}
\end{figure}

Figure \ref{fig:1}(a) shows the topography of SL NbS$_2$ on Au(111) measured by STM. Several domains of SL NbS$_2$ are visible with smaller regions of clean Au in between.  Although not visible in the STM image presented here, the dark regions between the  SL NbS$_2$ domains are identified as clean Au because either Au atoms and/or a typically distorted herringbone structure are observed \cite{Gronborg:2015aa}.  The SL NbS$_2$ domains cover approximately 75\,\% of the surface, and are irregular in shape with sharp boundaries between adjacent domains.  A well-ordered hexagonal moir\'{e}  is visible, due to the different periodicity of the substrate and the epitaxial layer.  The periodicity of the moir\'{e} is 20.7~{\AA}, as determined  from Fourier transform analysis of the STM data. The moir\'{e} pattern is qualitatively very similar to that observed for other epitaxial SL TMDCs on Au(111), particularly MoS$_2$, WS$_2$, and TaS$_2$ \cite{Gronborg:2015aa,Dendzik:2015aa,Sanders:2016aa}; and the minor distortions in the regularity of the moir\'{e}, especially close to edges and domain boundaries, can be understood in terms of local strain in the material  \cite{Zhang:2018ab}. 

Figure \ref{fig:1}(b) shows an atomically resolved image from which a hexagonal lattice constant of (3.4$\pm$0.3)~{\AA} was extracted.  This value is in agreement with the bulk lattice parameter of 3.3~{\AA} for 2H NbS$_2$ \cite{Wilson:1969aa}. It is not straightforward, on the basis of atomically-resolved STM images alone, to distinguish between the possible polymorphs of the SL structure, which are either trigonal prismatic (1H) or octahedral (1T),  since both polymorphs possess a hexagonal structure with presumably similar lattice constants. 

Figure \ref{fig:1}(c) shows a line profile taken across the edge of a domain boundary.   The profile reveals an apparent height difference of (3.1$\pm$0.2)~{\AA} between the top of the NbS$_2$ layer and the Au(111) surface. This value is consistent with the results for similar systems \cite{Gronborg:2015aa,Dendzik:2015aa,Sanders:2016aa}.
% for example single-layer WS$_2$ on Au(111) has an apparent height of (3.2$\pm$0.7)~{\AA},  confirming the interpretation of this structure as a single layer. 
 
\begin{figure}
	\begin{center}
		\includegraphics[width=1.0\linewidth]{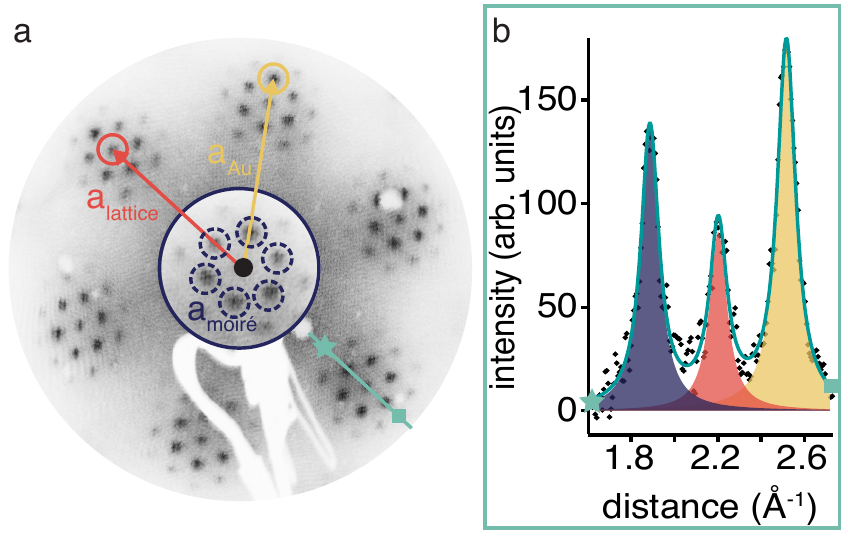}
		\caption{(a) LEED pattern for SL NbS$_2$ on Au(111) acquired at 30~K using an electron kinetic energy of 100~eV. The Au (yellow) and SL NbS$_2$ (red) reciprocal lattice vectors are shown. The LEED reveals six first-order spots in a hexagonal arrangement, and surrounding each of these individual spots are six intense satellite spots, also in a hexagonal arrangement.  The inset at the center of the LEED image is a magnified area around a first order spot. The satellite spots  arise from the moir\'{e} and are marked by the blue dashed lines.  (b) A line profile taken through the LEED image along the green line from the star marker to the square marker is presented. The intensity can be well described by three Lorentzian peaks, and the lattice parameter for SL NbS$_2$ was determined from the peak positions to be (3.29$\pm$0.03)~{\AA}. The fitting contributions are colour-coded according to the designation used for the LEED spots: SL NbS$_2$ (red), Au (yellow) and moir\'{e} (blue).} 
		\label{fig:2}
	\end{center}
\end{figure}

The overall ordering on the surface can be studied by LEED. Figure \ref{fig:2}(a) shows a LEED pattern revealing the integer order diffraction spots of both the NbS$_2$ (red circle) and the Au(111) lattice (yellow circle) and the additional spots due to the moir\'{e} (dashed blue circle). The two main lattices are well aligned and only one orientation of the SL NbS$_2$ is observed, albeit with the possibility of mirror domains \cite{Bana:2018aa}. An accurate determination of the SL NbS$_2$ lattice parameter can be achieved by fitting the peak positions in a line cut through the image, as shown in Figure \ref{fig:2}(b). The cut was taken along the green line from the star to the square. The known {2.88~\AA} lattice parameter of Au(111) was used as a calibration, and from the analyses of this and similar cuts in the corresponding directions, the lattice parameter of SL NbS$_2$ was determined to be (3.29$\pm$0.03)~{\AA}, in agreement with the STM results. We note that we do not observe any additional spots in the LEED that could hint at the existence of a charge density wave in SL NbS$_2$ at 30~K. 

\begin{figure}
	\begin{center}
		\includegraphics[width=1.0\linewidth]{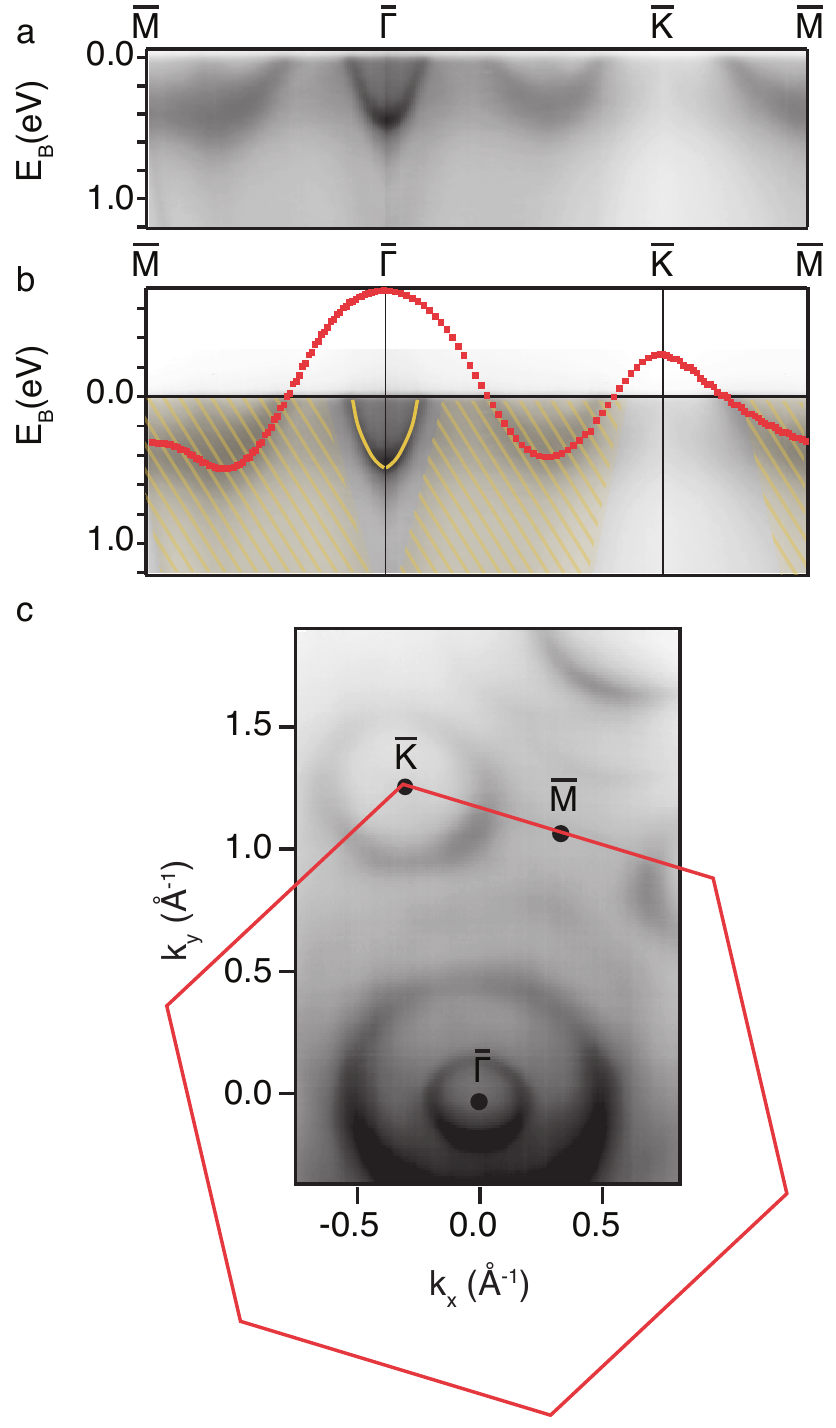}
		\caption{Photoemission intensity for SL NbS$_2$ on Au(111). (a) Photoemission intensity along the high symmetry lines of the NbS$_2$ 2D Brillouin zone acquired with a photon energy of 50\,eV. (b) The same ARPES data depicted in panel (a) but with the relevant features of the measured band structure marked: the Au surface state is highlighted by the yellow line; the projected bulk bands of the Au are demarcated by the yellow striped and shaded region; and, the calculated band structure for  SL NbS$_2$ reported by van Loon \textit{et al}. \cite{Loon:2018aa} is shown in red. (c) Photoemission intensity at the Fermi energy with $\bar{K}$, $\bar{M}$ and $\bar{\Gamma}$, indicated along with the NbS$_2$ 2D Brillouin zone overlaid in red.} 
		\label{fig:3}
	\end{center}
\end{figure}

Finally, we determine the electronic structure of SL NbS$_2$ using ARPES. The photoemission intensity along the high symmetry directions of the NbS$_2$ 2D Brillouin zone, acquired with 50\,eV photons, is shown in Figure \ref{fig:3}(a). The same data is reproduced in Figure \ref{fig:3}(b) but with salient features marked.  Some features crossing the Fermi level can be assigned to the well-known Au(111) surface state around $\bar{\Gamma}$ \cite{LaShell:1996aa} and the $sp$-bulk band of Au (very weak). The other visible features stem from SL NbS$_2$: a hole pocket around $\bar{\Gamma}$  and  a second hole pocket around $\bar{K}$($\bar{K^{\prime}}$). The parabolic Au(111) surface state is marked by the yellow line.  The yellow striped and shaded region indicates the projected bulk bands of Au(111). Overlaid on the data, in red, is the calculated bare dispersion for SL NbS$_2$ recently reported in Ref. \cite{Loon:2018aa}. From this ARPES data, we conclude that SL NbS$_2$ is metallic and that the general shape of the band structure is consistent with that expected for the 1H polytype \cite{Kuc:2011aa,Heil:2018aa}. The band structure is essentially the same  as observed for SL TaS$_2$ but without the strong spin-orbit splitting \cite{Sanders:2016aa}. 

The calculated band structure has been rigidly shifted by 0.15\,eV to higher binding energies in order to attain an approximate agreement between measurement and calculation; however, there is no rigid shift of the calculated band structure that can provide complete agreement with the measured data.   
%For the present shift, for instance, the observed occupied band width in the $\bar{M}\bar{\Gamma}$ and $\bar{\Gamma}$$\bar{K}$ directions appears to be smaller than the calculated value, whereas it is the other way around in the $\bar{K}$$\bar{M}$ direction. 
Perhaps the lack of quantitative agreement  is not entirely surprising, as almost the entire observed band structure falls with the projected bulk bands of Au(111) \cite{Takeuchi:1991aa} and therefore some hybridization with the substrate is expected to occur \cite{Bruix:2016aa,Dendzik:2017ab,Wehling:2016aa}. On the other hand, in the very similar case of SL TaS$_2$ on Au(111), an excellent agreement between calculated and observed bands can be achieved by a rigid energy shift \cite{Sanders:2016aa}. Lastly, we note that there are no indications of gap openings or back-folding of bands in Figure \ref{fig:3}(a), as might be expected in the presence of a CDW.

The photoemission intensity at the Fermi energy is shown in Figure \ref{fig:3}(c) with the high symmetry points indicated and the NbS$_2$ 2D Brillouin zone superimposed on the ARPES data in red.  From the size of the Fermi contour, the electron filling of the SL NbS$_2$ valence band can be estimated. Using the same approach as in Ref. \cite{Sanders:2016aa}, we determine that the upper valence band of SL NbS$_2$ on Au(111) is filled by 1.36$\pm$0.02 electrons per unit cell instead of the expected one electron. This corresponds to an electron concentration of (1.26$\pm$0.02)$\times 10^{15}$~cm$^{-2}$. Such strong doping effects have also been  observed for SL TaS$_2$ on Au(111) \cite{Sanders:2016aa}.  In that case, they are not necessarily caused by true doping of the SL material and can rather be explained by hybridization-induced ``pseudo-doping'' \cite{Wehling:2016aa}. This is also likely to be the case here. 

Despite the excellent crystalline quality revealed by STM and LEED, the SL NbS$_2$-related features observed in ARPES are very broad in contrast to the results from other other high-quality SL TMDCs. One possible reason for this is the hybridization between SL and substrate states \cite{Bruix:2016aa,Dendzik:2017ab}. Indeed, very sharp bands can only be observed in the case of semiconducting SL TMDCs on Au(111) or Ag(111) for the specific spectral region around the valence band maximum at $\bar{K}$  due to position of this band in a projected band gap of the substrate  \cite{Hinsche:2017aa,Mahatha:2019aa}. Here, the SL NbS$_2$ valence band maximum at $\bar{K}$ is above the Fermi energy, almost the entire observed band structure falls in the continuum of the Au bulk states, and is thus prone to hybridization. Another possible explanation for the broad bands is the many-body interactions discussed in Ref.  \cite{Loon:2018aa} that lead to a substantial broadening of the spectral function for the free-standing material. The calculated theoretical linewidth in Ref.  \cite{Loon:2018aa} is of the same order as the experimental values observed here. However, the additional interaction with the substrate makes a detailed analysis difficult. 

\section{Conclusions}

In summary, we have reported the epitaxial growth of SL NbS$_2$ on Au(111). Excellent structural quality can be realized, as revealed by STM and LEED. The resulting band structure is consistent with a 1H crystalline structure. It  reveals strong electron doping that could partly be due to hybridization with the substrate, as previously reported for SL TaS$_2$ \cite{Sanders:2016aa,Wehling:2016aa}. In contrast to the occupied band structure of SL TaS$_2$ on Au(111), however, clear deviations between the observed band structure and the rigidly shifted calculated bands for a free-standing layer are seen. Finally, the observed spectral features are rather broad. This could also be due to hybridization or to the predicted strong many-body interactions in this system \cite{Loon:2018aa}. This could perhaps be investigated further by transferring the growth procedure described here to a more inert substrate or, possibly, by intercalation-induced decoupling of the NbS$_2$ layer from the substrate \cite{Mahatha:2019aa}. 

\section{Acknowledgements}

We thank the authors of Ref. \cite{Loon:2018aa} for helpful discussions and for making their data available to us. 
This work was supported by the Danish Council for Independent Research, Natural Sciences under the Sapere
Aude program (Grants No. DFF-4002-00029 and DFF-6108-00409) and by VILLUM FONDEN via the Centre
of Excellence for Dirac Materials (Grant No. 11744) and the Aarhus University Research Foundation.  Affiliation with the Center for Integrated Materials Research (iMAT) at Aarhus University is gratefully acknowledged.

%\bibliography{NbS2bib}
\bibliographystyle{apsrev}

\end{document}